\documentstyle[12pt]{article}

\begin{document}

\newcommand{\be}{\begin{equation}}
\newcommand{\ee}{\end{equation}}
\newcommand{\no}{\nonumber}
\newcommand{\ba}{\begin{eqnarray}}
\newcommand{\ea}{\end{eqnarray}}
\newcommand{\la}{\langle}
\newcommand{\ra}{\rangle}

\begin{flushright}
hep-th/9612086 \\
preprint UT-763  \\
December, 1996\\
\end{flushright}

\bigskip

\bigskip

\begin{center}
{\Large \bf \it A New Description of the $E_6$ Singularity}
\end{center}

\bigskip

\bigskip

\begin{center}

Tohru Eguchi

\medskip

\medskip

{\it Department of Physics

\medskip

Faculty of Science

\medskip

University of Tokyo

\medskip

Tokyo 113, Japan}

\bigskip

and 

\bigskip

Sung-Kil Yang

\medskip

\medskip

{\it Institute of Physics

\medskip

University of Tsukuba

\medskip

Ibaraki 305, Japan}
\end{center}

\bigskip

\bigskip

\begin{abstract}
We discuss a new type of Landau-Ginzburg potential for the
$E_6$ singularity of the form 
$W=const+(Q_1(x)+P_1(x)\sqrt{P_2(x)})/x^3$
which featured in a recent study of heterotic/typeII
string duality. Here $Q_1,P_1$ and $P_2$ are polynomials of degree
15,10 and 10, respectively. We study the properties of the potential
in detail and show that it gives a new and consistent description of the
$E_6$ singularity.

\end{abstract}

\newpage

Recently Seiberg-Witten type construction of exact 
solutions for $N=2$ supersymmetric 
gauge theories has been generalized to the case of exceptional 
global and local gauge groups
by various authors \cite{MNa,LW,MNb,Ga,GMS}. 
$N=2$ theories of 
exceptional gauge symmetries are expected to arise from the 
degeneration of the $K_3$ surface in type II compactified string 
theory and play some basic role in typeII/heterotic duality.
On the other hand 
$SU(2)$ gauge theories with the $E_n$ global symmetry 
are expected to emerge on the world-volume of a $D$-brane probe
when the background $D$-branes are at a special geometrical 
configuration.
Since the local 
symmetry of the space-time string theory corresponds to the global 
symmetry of the world-volume theory on $D$-branes, theories with the 
exceptional local and global symmetries are expected to be closely
related to each other.

In ref.\cite{LW} by considering the degeneration of $K_3$ surface 
to an
$E_6$-type singularity Lerche and Warner obtained a spectral curve 
for the 
Seiberg-Witten theory of the $E_6$ gauge group,
\be
z+\Lambda^{24}/z+W_{E_6}(x;t_0,t_3,t_4,t_6,t_7,t_{10})=0.
\ee
Here the potential $W_{E_6}$ is given by
\be
W_{E_6}=-u_0(t_i)+{Q_1(x,t_i)\pm P_1(x,t_i)\sqrt{P_2(x,t_i)} \over x^3} 
\ee
with $Q_1(x),P_1(x)$ and $P_2(x)$ being certain polynomials in $x$ 
with degrees
$15,10$ and $10$, respectively.
A peculiar feature of the $E_6$ potential eq.(2) is the
appearance of the square root of the polynomial $P_2$. 

In this article we will study the 
potential (2) in some detail. We
shall show that it has a number of remarkable properties
and gives a new description of the $E_6$ singularity
and the topological field theory of $E_6$ type.

Let us recall that in the $SU(N)$ Seiberg-Witten theory the
spectral curve is given by \cite{Gor,MW}
\be
z+\Lambda^{2N}/z+2W_{SU(N)}(x,t_0,t_1,\cdots,t_{N-1})=0.
\ee
Here $W_{SU(N)}$ is an $N$-th order polynomial 
in $x$ and is the familiar Landau-Ginzburg superpotential for the 
$A_{N-1}$-type topological minimal model 
\be
A_{N-1}: \hskip3mm x^N+u_{N-2}x^{N-2}+u_{N-3}x^{N-3}+\cdots+u_0=0.
\ee
Substitution of $z=y-W_{SU(N)}$ into (3) leads to the 
Seiberg-Witten curve for the $SU(N)$ gauge theory \cite{AF,KLTY}
\be
y^2=W_{SU(N)}^2-\Lambda^{2N}.
\ee
Parameter $\{t_i\}$'s are related to $\{u_i\}$'s as
$u_i=t_i+\mbox{higher orders}$ and are the flat coordinates of the 
deformation 
of the $A_{N-1}$ singularity. 

Similarly the spectral curve for 
the $SO(2N)$ gauge theory is given by \cite{BL}
\be
z+\Lambda^{2(2N-2)}/z+W_{SO(2N)}(x,t_0,t_2,\cdots,t_{2(N-2)},t')=0.
\ee
$W_{SO(2N)}$ is again the Landau-Ginzburg potential of the 
$D_N$-type topological minimal model. It has the form
$W_{SO(2N)}=x^{2(N-1)}+t_{2(N-2)}x^{2(N-2)}+\cdots-t'^2/4x^2$  
which is obtained by eliminating the variable $y$ from the standard 
form of the $D_N$ singularity 
\ba &&D_N:
\hskip3mm x^{2(N-1)}+x^2y^2+u'y+u_{2(N-2)}x^{2(N-2)} \no \\
&&\hskip25mm +u_{2(N-3)}x^{2(N-3)}+\cdots +u_2x^2+u_0=0.
\ea 

In the following we would like to discuss the relation of the 
potential eq.(2) and the standard form of the $E_6$ singularity
\be
E_6: \hskip3mm x^4+y^3+u_{10}x^2y+u_7xy+u_6x^2+u_4y+u_3x+u_0=0.
\label{E6}
\ee
We shall show that the topological field theory described by the 
potential eq.(2) is equivalent to that of the conventional $E_6$ singularity
theory (\ref{E6}). In particular the free-energies of the 
theories exactly agree with each other. Thus the superpotential eq.(2) 
may be regarded as a novel single-variable description of the $E_6$ 
singularity. We expect that it gives rise 
to a new type of 
integrable hierarchy as is the case of the well-known KP hierarchy 
which arises 
>from the theory of $A-D$ singularities.

Let us first write down explicitly the polynomials $Q_1,P_1,P_2$ \cite{LW},
\ba
Q_1(x;t_i)&\hskip-3mm =& \hskip-3mm 270x^{15}+(171+57\sqrt{3})t_{10}x^{13} 
+(54+27\sqrt{3})t_{10}^{2}x^{11} \no \\
& &\hskip-15mm+
(126+84\sqrt{3})t_7x^{10}+\Big(({35 \over 4}+{175 \over 36}\sqrt{3})t_{10}^{3}
+(144+72\sqrt{3})t_6\Big)x^{9} \no \\
& &\hskip-15mm
+({135 \over 2}+{81 \over 2}\sqrt{3})t_{10}t_7x^{8} 
+\Big(({225 \over 4}+{125 \over 4}\sqrt{3})t_6t_{10}
+({245 \over 384}+{35 \over 96}\sqrt{3})t_{10}^{4} 
\no \\
& &\hskip-15mm
+(135+81\sqrt{3})t_4\Big)x^{7}
+\Big((126+72\sqrt{3})t_3+(10+{35 \over 6}\sqrt{3})t_{10}^{2}t_7\Big)
x^{6} \no \\
& &\hskip-15mm
+\Big(({63 \over 4}+9\sqrt{3})t_7^{2}+(36+21\sqrt{3})t_{10}t_4 
+({11 \over 768}+{19 \over 2304}\sqrt{3})t_{10}^{5} \no \\
& &\hskip-15mm
+({21 \over 4}+3\sqrt{3})t_{10}^{2}t_6\Big)x^{5} \no \\
& &\hskip-15mm
+\Big(({33 \over 2}+{19 \over 2}\sqrt{3})t_{10}t_3
+({19 \over 48}+{11 \over 48}\sqrt{3})t_{10}^{3}t_7 
+(24+14\sqrt{3})t_7t_6\Big)x^4 \no \\
& &\hskip-15mm
+(-{11 \over 8}-{19 \over 24}\sqrt {3})t_{10}t_7^{2}x^{3}
+({45 \over 4}+{13 \over 2}\sqrt{3})t_7t_3x+ 
({5 \over 8}+{13 \over 36}\sqrt{3})t_7^{3}, \no \\
& & \\
P_1(x;t_i)& \hskip-3mm =& \hskip-3mm 78x^{10}+(30+10\sqrt{3})t_{10}x^{8}+
({14 \over 3}+{7 \over 3}\sqrt{3})t_{10}^{2}x^{6}
+({33 \over 2}+11\sqrt{3})t_7x^{5} \no \\
& &+\Big(({1 \over 4}+{5 \over 36}\sqrt{3})t_{10}^{3}+
(16+8\sqrt{3})t_6\Big)x^{4}
+({25 \over 12}+{5 \over 4}\sqrt{3})t_{10}t_7x^{3} \no \\
& &+\Big((5 + 3\sqrt{3})t_4+({7 \over 3456}+{1 \over 864}\sqrt{3})t_{10}^{4}
+({3 \over 4}+{5 \over 12}\sqrt{3})t_6t_{10}\Big)x^{2}
\no \\
& &+(-{7 \over 2}-2\sqrt{3})t_3x 
+(-{7 \over 12}-{1 \over 3}\sqrt{3})t_7^{2}, \no \\
& &     \\
P_2(x;t_i)& \hskip-3mm =& \hskip-3mm 12x^{10}+(6+2\sqrt{3})t_{10}x^{8}+
({4 \over 3}+{2 \over 3}\sqrt{3})t_{10}^{2}x^{6}+(6+4\sqrt{3})t_7x^{5} \no \\
& &+\Big((8+4\sqrt {3})t_6 + 
({1 \over 8}+{5 \over 72}\sqrt{3})t_{10}^{3}\Big)x^{4}
+({5 \over 3}+\sqrt{3})t_{10}t_7x^{3} \no \\
& &+\Big((10+6\sqrt{3})t_4 + 
({7 \over 1728}+{1 \over 432}\sqrt{3})t_{10}^{4}
+({3 \over 2}+{5 \over 6}\sqrt{3})t_{10}t_6\Big)x^2 \no \\
& &+(14+8\sqrt{3})t_3x
+({7 \over 12}+{1 \over 3}\sqrt{3})t_7^{2}.
\ea
(Our parametrizations are different from those of \cite{LW}.
Relation between our $t_i$'s and $w_i$'s of ref.\cite{LW}
is given by 
\ba
&&w_1={1 \over 12}(3+\sqrt{3})t_{10},
w_2={1 \over 24}(3+2\sqrt{3})t_7, \no \\
&&w_3={1 \over 24}(2+\sqrt{3})t_6
-{1 \over 6912}(9+5\sqrt{3})t_{10}^3, \no \\
&&w_4=-{1 \over 96}(5+3\sqrt{3})t_4
-{1 \over 331776}(7+4\sqrt{3})t_{10}^4+{1 \over 1152}(9+5\sqrt{3})t_6t_{10},
\no \\
&&w_5=-{1 \over 96}(7+4\sqrt{3})t_3-{1 \over 6912}(12+7\sqrt{3})t_7t_{10}^2,
 \\
&&w_6={1 \over 55296}(19+11\sqrt{3})t_4t_{10}^2
-{1 \over 1990656}(33+19\sqrt{3})t_6t_{10}^3 \no \\
&&+{1 \over 4608}(7+4\sqrt{3})t_6^2
-{1 \over 55296}(33+19\sqrt{3})t_7^2 t_{10}-{1 \over 576}(45+26\sqrt{3})t_0.)
\no
\ea
 
Note that the polynomials $P_1,P_2$ are related to each other by an 
identity
\be
P_1=-P_2+{3 \over 4}x{dP_2 \over dx}.
\ee 

For the sake of the definiteness let us 
choose the $+$ sign in eq.(2). We
also introduce a normalization factor so that the $E_6$ potential is
now given by 
\be
W_{E_6}={1 \over 270+156\sqrt{3}}
\Big(-u_0+{Q_1+P_1\sqrt{P_2} \over x^3}\Big).
\ee
Here $u_0$ is expressed in terms of $\{t_i\}$'s as
\ba
u_0& \hskip-3mm =& \hskip-3mm -(270+156\sqrt{3})t_0-{1 \over 16}(19+11\sqrt{3})
t_{10}^2t_4-{1 \over 576}(33+19\sqrt{3})t_{10}^3t_6 \no \\
&&-{1 \over 4}(21+12\sqrt{3})t_6^2
-{1 \over 16}(33+19\sqrt{3})t_{10}t_7^2.
\ea
Expansion of $W_{E_6}$ at $x=\infty$ yields
\ba
W_{E_6}& \hskip-3mm =& \hskip-3mm 
x^{12}+t_{10}x^{10}+{3 \over 8}t_{10}^2x^8+t_7x^7+(t_6+{7 \over 108}t_{10}^3)
x^6+{1 \over 2}t_{10}t_7x^5 \no \\
&&+(t_4+{5 \over 12}t_{10}t_6+{35 \over 6912}t_{10}^4)x^4
+(t_3+{5 \over 72}t_{10}^2t_7)x^3
 \no \\
&&+({1 \over 8}t_7^2+{1 \over 24}t_{10}^2t_6+{1 \over 4}t_{10}t_4
+{1 \over 6912}t_{10}^5)x^2 
+({1 \over 6}t_7t_6+{1 \over 6}t_{10}t_3+{1 \over 432}t_{10}^3t_7)x 
\no \\
&&+t_0+{1 \over 24}t_6^2+{1 \over 144}t_{10}^2t_4+
{1 \over 144}t_{10}t_7^2+{1 \over 1728}t_{10}^3t_6
+{1 \over 1492992}t_{10}^6 \no \\
&&+\Big({1 \over 12}(1-\sqrt{3})t_6t_4+
{1 \over 24}(1-\sqrt{3})t_7t_3\Big)x^{-2}+\cdots \hskip1mm . \no \\
&&
\ea
The precise dependence of the polynomials $Q_1,P_1,P_2$ and $u_0$ 
on the flat coordinates 
$t_i \hskip1mm (i=0,3,4,6,7,10)$ has been determined by requiring 
the following relations on the residue integrals
\ba
&&t_{10-i}={12 \over i+1}\oint dx W_{E_6}(x;t_j)^{i+1 \over 12}, \hskip2mm 
i \in I_{E_6}\equiv \{0,3,4,6,7,10\}, \\
&&0=\oint dx W_{E_6}(x;t_j)^{i+1 \over 12}, \hskip22mm i \not\in I_{E_6}
\ea
where the integral is around $x=\infty$. 

We note that the relations eq.(17),(18) are analogues of those 
well-known in the
singularity theory of $A_{N-1}$-type \cite{DVV}
\be
t_{N-2-i}={1 \over i+1}\oint dx W_{A_{N-1}}(x;t_i)^{{i+1 \over N}}, 
\hskip3mm i\in I_{A_{N-1}}\equiv\{0,1,2,\cdots,N-2\}.
\ee
When $i=N-1$, the right-hand-side of (19) vanishes automatically since
$W_{A_{N-1}}$ is a polynomial in $x$. 

Further relations are known to hold for one-point functions 
\cite{EKYY} 
\ba
\la \sigma_n(\phi_i) \ra
& \hskip-3mm =&\hskip-3mm c_{n,i}^N \oint dx 
W_{A_{N-1}}(x;t_i)^{{i+1 \over N}+n+1}, 
\hskip1mm i\in I_{A_{N-1}}, 
\hskip1mm n=0,1,2,\cdots \no \\
&& \\
&& \hskip-20mm c_{n,i}^N=\Pi_{m=0}^{n+1}(i+1+Nm)^{-1}.
\ea
Here $\phi_i$ denotes the primary field
$\phi_i=\partial W_{A_{N-1}}/\partial t_i$ 
and $\sigma_n(\phi_i)$ is the $n$-th gravitational 
descendant of 
$\phi_i$. (20) gives descendant one-point functions 
(or two-point functions $\la P\sigma_{n+1}(\phi_i) \ra$ with the 
puncture operator $P$) and thus equals the Gelfand-Dikii potentials
of the KP hierarchy in the dispersionless limit.

In our case primary fields are again defined by the
derivatives of the potential 
\be
\phi_i={\partial W_{E_6} \over \partial t_i},
\ee
and one-point functions of their descendants are given by
\ba
&&\la \sigma_n(\phi_i)\ra =12^{n+2}\hskip1mm c_{n,i}^{12}
\hskip1mm \eta_{i,10-i} \oint dx 
W_{E_6}(x;t_i)^{{i+1 \over 12}+n+1}, \hskip3mm i \in I_{E_6} \no \\
&&            \\
&&\oint dx W_{E_6}(x;t_i)^{{i+1 \over 12}+n}=0, \hskip10mm i\not\in I_{E_6}
\hskip10mm n=0,1,2,\cdots.
\ea
Here $\eta_{ij}$ is the topological metric defined by 
\be
\eta_{ij}={12^2 \over \hskip-1mm 13}\partial_i\partial_j
\oint dx W_{E_6}^{13/12}=\langle \phi_i\phi_jP\rangle.
\ee
Its non-vanishing elements are given by $\eta_{0,10}=1,
\eta_{3,7}=(3-\sqrt{3})/2,\eta_{4,6}=2-\sqrt{3}$.
By direct computation we can check that the above vanishing
conditions (24) hold with our $E_6$ potential eq.(14).
Eq.(23) gives the Gelfand-Dikii potentials of the $E_6$ theory.

 It is instructive to compare the $E_6$ superpotential 
with that of the $SU(12)$ 
theory. $E_6$ and $SU(12)$ have the same Coxeter number $12$ 
and hence their potentials have the same leading term $x^{12}$. 
Recall that the Landau-Ginzburg potential of $A_{N-1}$
theory is given by the determinant \cite{DVV}
\be
{1 \over N}{d W_{A_{N-1}} \over dx}(x,t_i')=\left |\begin{array}{cccccc}
x & -1 & 0& 0 & 0& 0 \\
t_{N-2}' & x & -1 & 0 & 0 &0 \\
t_{N-3}' & t_{N-2}' & x & -1 & 0 & 0 \\
.. & .. & .. & ..& ..& ..  \\
t_2' & .. & .. & ..& x& -1  \\
t_1' & t_2' & .. & t_{N-3}' & t_{N-2}'& x \end{array}  \right |.
\ee
By expanding the determinant and setting the parameters
$t_1',t_2',t_5',t_8',t_9'$ to zero one finds that $W_{A_{11}}$ reproduces 
the "positive" 
part of the $E_6$ potential,
\ba
&&W_{A_{11}}(x;t_0',t_3',t_4',t_6',t_7',t_{10}',t_1'=t_2'=t_5'=t_8'=t_9'=0) 
\no \\
&&=W_{E_6}(x;t_0,t_3,t_4,t_6,t_7,t_{10})|_+, \hskip2mm t_i=12t_i', \hskip1mm
i\in I_{E_6}
\ea
($|_+$ denotes the sum of terms with non-negative powers of 
$x$). 
It seems likely that one generates the $E_6$ potential by 
starting from 
that of $SU(12)$ and demanding the vanishing conditions (24) which would 
uniquely fix the negative part of $W_{E_6}$. 
We note that the positive parts of the primary fields of the $E_6$ theory 
$\phi_i|_+ \hskip1mm (i\in I_{E_6})$ coincide with the primary fields 
$\phi_i \hskip1mm (i\in I_{E_6}\subset I_{SU(12)})$ 
of $SU(12)$ theory since primary fields are 
defined by the derivatives of the potentials.

Let us next look at the closure of the operator product expansion (OPE)
in the $E_6$ theory.
In the case of the $A_{N-1}$ theory the closure of the operator
algebra is more or less obvious since the set of the exponents 
$I_{A_{N-1}}=0,1,2,\cdots,N-2$
does not have "missing" degrees except for $i=N-1$ which is
the degree of $dW/dx$ and interpreted as the BRS trivial operator.
On the other hand in the case of $E_6$ the exponents $I_{E_6}=0,3,4,6,7,10$
have several missing degrees and it is not obvious how the closure of
OPE can be achieved. 

Now we would like to show that OPE nevertheless
holds with the $E_6$ Landau-Ginzburg potential eq.(14) and leads 
to the structure 
constants (3-point functions) which agree completely with those 
of the standard $E_6$ theory of eq.(8).

We first recall the general structure of the operator product 
expansion in the Landau-Ginzburg theory,
\be
\phi_i(x)\phi_j(x)=Q_{ij}(x)W(x)'+
\sum_{\ell \in I}c_{ij}^{\hskip3mm\ell}\phi_{\ell}(x), \hskip5mm 
i,j \in I.
\ee
Here the functions $Q_{ij}$ (contact terms) are defined by
\be
{\partial \phi_i \over \partial t_j}={\partial Q_{ij} \over \partial x}.
\ee
In the case of the $A_{N-1}$ theory it is easy to show that
\be
Q_{ij}(x)=\phi_{i+j+1-N}(x)
\ee
and $Q_{ij}(x)$ vanishes when $i+j<N-1$. 

In the $E_6$ case, however, $Q_{ij}(x)$ 
does not vanish for any value of $i,j$ and behaves as $Q_{ij}(x)
\approx x^{i+j+1-N}$. When one considers the operator 
product $\phi_i(x)
\phi_j(x)$ with $i+j \not \in I_{E_6}$, the leading term $x^{i+j}$
is canceled by the BRS-exact piece in the right-hand-side of
(28) 
and the remaining terms can be written as the sum over primary fields. 
This is the mechanism of how the OPE works in the $E_6$ theory.
Direct check of eq.(28) is non-trivial since both 
sides of OPE become infinite series when expanded around $x=\infty$.

Using the algebraic computation software (Maple) 
we have verified the above OPE relations
directly  by substituting the expressions for $W$, 
$\phi_i={\partial W \over \partial t_i}$ and 
$Q_{ij}=\int dx {\partial^2 W \over \partial t_i \partial t_j}$ 
into eq.(28) without expanding the square-root of $P_2$.
($x^0$-term of $Q_{ij}$ is given by 
1/12 of $x^0$-term of $\phi_{i+j-11}^{SU(12)}$
where $\phi_j^{SU(12)}$ is the primary field of $SU(12)$ theory
at $t_i'=t_i/12, \hskip1mm i\in I_{E_6}$ and $t_i'=0, 
\hskip1mm i\not\in I_{E_6}$).
At the same time we have determined the structure constants 
$\{c_{ij}^{\hskip3mm\ell}\}$ and verified that they agree with the
3-point functions $\eta_{k\ell}c_{ij}^{\hskip2mm \ell}=c_{ijk}
=\partial_i\partial_j\langle\phi_k\rangle$.

Checking OPE is much simplified if we introduce a new 
representation of primary fields,
\ba
&&\phi_i(x;t_j)={\partial A_i(x;t_j) \over \partial x}, 
\hskip10mm i\in I_{E_6} \\
&&A_i(x;t_j)=R_i(x;t_j)+S_i(x;t_j)\sqrt{P_2(x;t_j)}. 
\ea
If we use the identity eq.(13), it is easy to derive
\ba
&&R_i(x;t_j)={\partial Q_2(x;t_j) \over \partial t_i},\hskip4mm 
Q_2(x;t_j)=\int^{x} (-u_0(t_j)+{Q_1(x;t_j) \over x^3})dx, \no \\
&&S_i(x;t_j)={3 \over 4x^2}{\partial P_2(x;t_j) \over  \partial t_i}.
\ea
OPE relations reduce to several identities obeyed by the functions 
$R_i,S_i$.

Let us now turn to the discussions of the structure constants 
$\{c_{ij}^{\hskip3mm\ell}\}$. As is obvious from the definition eq.(28),
$c_{ij}^{\hskip3mm\ell}$ 
are symmetric in the indices ${ij}$, however, not symmetric in the
third index $\ell$. 
If one wants to bring them into a fully symmetric form,
we have to introduce rescaling factors $a_i$ which convert our
flat coordinates into those of the conventional $E_6$ theory eq.(8).
One finds that 
\ba
&&t_i=a_iT_i, \hskip15mm i\in I, \\
&&C_{ij}^{\hskip3mm \ell}={a_ia_j \over 
a_{\ell}a_0}c_{ij}^{\hskip3mm \ell},
\ea
where $T_i$'s and $C_{ij}^{\hskip3mm \ell}$'s are the flat coordinates and
structure constants of the 
conventional $E_6$ theory \cite{KST} (indices of 
$C_{ij}^{\hskip3mm \ell}$ 
are raised and lowered by the metric $\delta_{i+j,10}$).
Values of $a_i$'s are given by
\ba
&&a_0=192(-45+26\sqrt{3}), \hskip7mm a_3=96\sqrt{2}(7-4\sqrt{3}), \no \\
&&a_4=48\cdot 3^{1/3}(-5+3\sqrt{3}), \hskip3mm
a_6=48(-2+\sqrt{3}),  \no \\
&&a_7=8\sqrt{2}\cdot 3^{1/3}
(3-2\sqrt{3}), \hskip3mm a_{10}=-4\cdot 3^{1/3}(3-\sqrt{3}).  \\
\ea
$C_{ij}^{\hskip3mm \ell}$ can be read off from the formula of
the free-energy \cite{KST}
\ba
&&F_{E_6}=T_0T_4T_6+T_0T_3T_7+{T_0^2T_{10} \over 2}+{T_3^2T_{4} \over 2}
+{T_3T_6^2T_{7} \over 2}+{T_4^2T_{7}^2 \over 4} \no \\
&&+{T_6T_7^4 \over 12}+{T_4^3T_{10} \over 6}+{T_3^2T_6T_{10} \over 2}
+{T_6^4T_{10} \over 12}+{T_4T_6T_7^2T_{10} \over 2}
+{T_3T_7^3T_{10} \over 6} \no \\
&&+{T_3T_4T_7T_{10}^2 \over 2}+{T_6^2T_7^2T_{10}^2 \over 4}
+{T_4T_6^2T_{10}^3 \over 6}+{T_3T_6T_7T_{10}^3 \over 6}
+{T_7^4T_{10}^3 \over 24}+{T_3^2T_{10}^4 \over 24} \no \\
&&+{T_4T_7^2T_{10}^4 \over 24}+{T_4^2T_{10}^5 \over 60}
+{T_6 T_7^2T_{10}^5 \over 24}+{T_6^2T_{10}^7 \over 252}
+{T_7^2T_{10}^8 \over 3^22^6}+{T_{10}^{13} \over 11\cdot13\cdot2^43^4} 
\hskip2mm.\no \\
&&
\ea

Thus we have established that the single-variable potential (14)
describes correctly the $E_6$ singularity although the appearance of
the square root of a polynomial looks odd at first sight.
In fact one may consider other examples of Landau-Ginzburg potentials
possessing a square root of polynomials. If one considers, for example, 
the spectral
curve of $SU(5)$ theory obtained from the Toda system based on the
anti-symmetric representation (10-dim. representation) \cite{MW}, 
one finds
\be
z+{\Lambda^{10} \over z} +W_{SU(5)}(x;t_i)=0
\ee
with
\be
W_{SU(5)}(x;t_i)={2 \over 11+5\sqrt{5}}
\Big(-u_0+{q_1(x)+p_1(x)\sqrt{p_2(x)} \over 2}\Big).
\ee
Polynomials $q_1,p_1,p_2$ are given by
\ba
&&q_1(x)=11x^5+(6+2\sqrt{5})t_3x^3+
({7 \over 2}+{7 \over 2}\sqrt{5})t_2x^2 \no \\
&&\hskip10mm+(({7 \over 10}+{3 \over 10}\sqrt{5})t_3^2+(6+2\sqrt{5})t_1)x
+(2+\sqrt{5})t_3t_2  \\
&& \no \\
&&p_1(x)=5x^3+
({3 \over 2}+{1\over 2}\sqrt{5})t_3x+
({1 \over 2}+{1 \over 2}\sqrt{5})t_2 \\
&& \no \\
&&p_2(x)=5x^4+(3+\sqrt{5})t_3x^2+(2+2\sqrt{5})t_2x \no \\
&&\hskip20mm +({7 \over 10}+{3 \over 10}\sqrt{5})t_3^2
+(6+2\sqrt{5})t_1
\ea
and 
\be
u_0=-{11+5\sqrt{5} \over 2}t_0+{2+\sqrt{5} \over 5}t_2t_3.
\ee
It is easy to check that this system is equivalent to the
standard $SU(5)$ theory described by the superpotential eq.(26) with $N=5$.
Analysis goes completely  parallel to the case of the $E_6$ theory and one
finds in the end that the free-energies of the theories (39), (26) coincide.

We recall that by quotienting $E_6$ by the diagram automorphism
one obtains the Lie algebra $F_4$. The diagram automorphism acts on the 
parameters as $t_3,t_7 \rightarrow -t_3,-t_7$ while other parameters are left
unchanged \cite{Zu}. Thus we conjecture that the spectral curve for
the $N=2$ gauge theory with $F_4$ gauge group is given by the superpotential
(2) with $t_3=t_7=0$.

In this paper we have seen that the spectral curve describing the 
Seiberg-Witten theory for gauge group $G$ has the form
\be
z+{\Lambda^{2g^*} \over z}+W_G(x;t_i)=0
\ee
where $W_G$ is a single-variable Landau-Ginzburg potential for the singularity
of $G$ type ($g^*$ is the dual Coxeter number of $G$). It is known that 
actually the term $z+\Lambda^{2g^*}/z$ 
is also the Landau-Ginzburg potential for the $\sigma$-model with $CP^1$ 
target space \cite{EY}. Thus (44) gives a sum of potentials for the $CP^1$ and
the ALE spaces and therefore the $K_3$-fibered Calabi-Yau manifolds at the
degeneration limits. It is quite interesting to see if one can develop
a precise connection between the $N=2$ gauge theory and the
topological $\sigma$-model with the Calabi-Yau target space.

\bigskip

\bigskip

Research of T.E. and S.K.Y are supported in part by the 
Grant-in-Aid for Scientific
Research on Priority Area 213 "Infinite Analysis", Japan Ministry of
Education.


\newpage

\begin{thebibliography}{100}

\bibitem{MNa}
J.A. Minahan and D. Nemeshansky {\it An $N=2$ Superconformal Fixed 
Point with Global $E_6$ Symmetry}, hep-th/9608047.

\bibitem{LW}  
W. Lerche and N.P. Warner {\it Exceptional SW Geometry from ALE 
Fibrations}, hep-th/9608183.

\bibitem{MNb}
J.A. Minahan and D. Nemeshansky {\it Superconformal fixed Points 
with $E_n$ Global Symmetry}, hep-th/9610076.

\bibitem{Ga}
O.J. Ganor, {\it Toroidal Compactification of 6D Non-critical
Strings down to 4 Dimensions}, hep-th/9608109.

\bibitem{GMS}
O.J. Ganor, D. Morrison and N. Seiberg, {\it Branes, Calabi-Yau 
Spaces, and Toroidal Compactification of the $N=1$ Six-Dimensional 
$E_8$ Theory}.

\bibitem{Gor}
A. Gorsky, I. Krichever, A. Marshakov, A. Mironov and A. Morozof,
Phys. Lett. {\bf B355} (1995) 466.

\bibitem{MW}
E.J. Martinec and N.P. Warner, Nucl. Phys. {\bf B459} (1996) 97.


\bibitem{AF}
P. Argyres and A. Faraggi, Phys. Rev. Lett. {\bf 74} (1995) 3931.

\bibitem{KLTY}
A. Klemm, W. Lerche, S. Theisen and S. Yankielowicz, 
Phys. Lett. {\bf B344} (1995) 169.

\bibitem{BL}
A. Brandhuber and K. Landsteiner, Phys. Lett. {\bf B358} (1995) 73.

\bibitem{DVV}
R. Dijkgraaf, E. Verlinde and H. Verlinde,
Nucl. Phys. {\bf B352} (1991) 59.

\bibitem{EKYY}
T. Eguchi, H. Kanno, Y. Yamada and S.-K. Yang,
Phys. Lett. {\bf 305B} (1993) 235; 
T. Eguchi, Y. Yamada and S.-K. Yang,
Mod. Phys. Lett. {\bf A8} (1993) 1627.

\bibitem{KST}
A. Klemm, S. Theisen and M.G. Schmidt, Int. J. Mod. Phys.
{\bf A7} (1992) 6515.

\bibitem{Zu}
J.-B. Zuber, Mod. Phys. Lett. {\bf A9} (1994) 749.

\bibitem{EY} T. Eguchi and S.-K. Yang,
Mod. Phys. Lett. {\bf A9} (1994) 2893;
T. Eguchi, K. Hori and S.-K. Yang,
Int. J. Mod. Phys. {\bf A10} (1995) 4203.





\end{thebibliography}
\end{document}